\documentclass{nature}
\usepackage{epsfig}
\usepackage{amsmath,mathrsfs,amsbsy,color,graphicx,bm,amsthm,amsfonts}
\usepackage{units}
\usepackage{bbm}
\usepackage{times}
\usepackage{dcolumn}
\usepackage{mathrsfs}
\usepackage{amssymb,float}
\begin{document}
\baselineskip=0.5cm
\title{Quantum coherence and correlations in quantum system}
\author{Zhengjun Xi$^{1,\star}$, Yongming Li$^{1}$, Heng Fan$^{2,3}$}
\maketitle

\begin{affiliations}
\item
College of Computer Science, Shaanxi Normal University, Xi'an, 710062,
P. R. China
\item
Beijing National Laboratory for Condensed Matter Physics, Institute
of Physics, Chinese Academy of Sciences, Beijing, 100190, P. R. China.

\item Collaborative Innovation Center of Quantum Matter, Beijing, P. R. China

$^\star$e-mail:xizhengjun@snnu.edu.cn

\end{affiliations}

\begin{abstract}
\baselineskip=0.5cm
Criteria of measure quantifying quantum coherence, a unique property of quantum system, are proposed recently.
In this paper, we first give an uncertainty-like expression relating the coherence and the entropy of quantum system. This finding allows us to discuss the relations between the entanglement and the coherence. Further, we discuss in detail the relations among the coherence, the discord and the deficit in the bipartite quantum system. We show that, the one-way quantum deficit is equal to the sum between quantum discord and the relative entropy of coherence of measured subsystem.
\end{abstract}

Quantum coherence arising from quantum superposition plays a central role in
quantum mechanics. Quantum coherence is a common necessary condition for
both entanglement and other types of quantum correlations, and it is also an important physical resource in
quantum computation and quantum information processing. Recently, a
rigorous framework to quantify coherence has
been proposed\cite{Baumgratz13} (or see early work\cite{Aberg06}). Within such a framework for the coherence,
one can define suitable measures, include the relative entropy and the $l_1$-
norm of coherence\cite{Baumgratz13}, and a measure
by the Wigner-Yanase-Dyson skew information\cite{Girolami14}. Quantum coherence has received a lot of attentions\cite{Angelo13, Marvian13,Rosario13,Levi14,Marvian14,Lostaglio14,Hai14,Monras14,Karpat14, Aberg14,Xi14}. We know that quantum coherence and the entanglement are related to quantum superposition, but we are not sure of the exact relations between quantum coherence and the entanglement, is there a quantitative relation between the two of them?

On the other hand, it is well known that entanglement does not account for all nonclassical correlations (or quantum correlations)
 and that even correlation of separable state does not completely be classical. Quantum discord \cite{Ollivier02, Henderson01} and quantum deficit \cite{Oppenheim02} have been viewed as two possible quantifiers for quantum correlations. There have been much interest in characterizing and interpreting their applications in quantum information processing\cite{horodeckirmp,Modi12,Chuan12,Streltsov12,Amico-RMP,Cramer-RMP,Cui-Gu,Haldane-PRL,FanLiuPRL,
FanWangPRL,FanCuiPRX, Zwolak12,Pirandola14}. In particular, Horodecki \emph{et al.} \cite{Horodecki05} discussed the relationship between the discord and quantum deficit in the bipartite quantum system. If only one-way classical communication from one party to another is allowed,  they showed that the one-way quantum deficit is an upper bound of quantum discord via the local von Neumman measurements on the party.
Curiously, up to now, no attempt for a transformed framework between them has been reported. In other word, is there a more clear quantitative relations between them?

  %
In the present work, we will resolve the above questions via quantum coherence.
We only focus on particular the entropic form, also called relative entropy of coherence,
which enjoys the properties of physical interpretation
and being easily computable\cite{Baumgratz13}. Firstly, we derive an uncertainty-like expression which states that the sum of the coherence and the entropy in quantum system is bounded from the above by $\log_2d$, where $d$ is the dimension of the quantum system.  As an application, we discuss the relations between the entanglement and the coherence. Meanwhile, we find that the relative entropy of coherence satisfies the super-additivity. In the bipartite quantum system, based on the projective measurement in which the relative entropy of coherence is quantified, we obtain that the increased entropy produced by the local projective measurement is equal to the sum between the quantum correlation destroyed by this measurement and the relative entropy of coherence of the measured subsystem.
Since the incoherent states under two different bases are unitarily equivalent, then there are same matrix elements under the different bases for given quantum state. These two facts are the reasons that we study in detail the explicit expressions of the discord and the deficit in terms of the relative entropy of coherence in the bipartite quantum system. In this way, we can give a clear quantitative relation between the discord and the deficit.

\vspace{.8cm}
\noindent{\large\bf Results}

\vspace{.1cm}
\noindent
{\bf Measure of quantum coherence.}
Consider a finite-dimensional Hilbert space $\mathcal{H}$ with $d=dim({\mathcal{H}})$. Fix a basis $\{|i\rangle\}_{i=1}^{d}$,
we take the suggestion given by Baumgratz \emph{et al.}\cite{Baumgratz13}, let $I$ be a set of the incoherent states, which is of the form
\begin{equation}
\sigma=\sum_{i=1}^d\sigma_i|i\rangle\langle i|,
\end{equation}
where $\sigma_i\in[0,1]$ and $\sum_i\sigma_i=1$.
Baumgratz \emph{et al.} proposed that any proper measure of the coherence $C$ must satisfy the following conditions:
\begin{itemize}
\item [{(C1)}]  $C(\rho)\geq 0$ for all quantum states $\rho$, and $C(\rho)=0$ if and only if $\rho\in \mathcal{I}$.
\item [{(C2a)}]  Monotonicity under incoherent completely positive
and trace preserving maps (ICPTP) $\Phi$, i.e.,
$C(\rho) \geq C(\Phi(\rho))$.
\item [{(C2b)}]  Monotonicity for average coherence under subselection based on measurements outcomes: $C(\rho)\geq \sum_n p_n C(\rho_n) $, where $\rho_n=K_n\rho K_n^\dag/p_n$ and $p_n=\mathrm{Tr}(K_n \rho K_n^\dag)$ for all $\{K_n\}$ with $\sum_n K_n^{\dagger}K_n= I$ and $K_n \mathcal{I} K_n^\dagger\subseteq\mathcal{I}$.
\item [{(C3)}]  Non-increasing under mixing of quantum states (convexity), i.e.,
$\sum_ip_i C(\rho_i)\geq C(\sum_ip_i\rho_i),$
 for any ensemble $\{p_i,\rho_i\}$.
\end{itemize}
Note that conditions (C2b) and (C3) automatically
imply condition (C2a). We know that the condition (C2b) is important as it allows for sub-selection
based on measurement outcomes, a process available in well controlled quantum experiments\cite{Baumgratz13}. It has been shown that the relative entropy and $l_1$-norm satisfy all conditions. However, the measure of coherence induced by the squared Hilbert-Schmidt norm satisfies conditions (C1), (C2a), (C3), but not (C2b). Recently, we also find that the measure of coherence induced by the fidelity
does not satisfy condition (C2b), and an explicit example is presented~\cite{Xi14}.

In the following, we only consider the measure of relative entropy of coherence. For any quantum state $\rho$ on the Hilbert space $\mathcal{H}$, the relative entropy of coherence\cite{Baumgratz13} is defined as
\begin{equation}
C_{\mathrm{RE}}(\rho):=\min_{\sigma\in I}S(\rho||\sigma),
\end{equation}
where $S(\rho||\sigma)=\mathrm{Tr}(\rho\log_2\rho-\rho\log_2\sigma)$ is the relative entropy\cite{Nielsen}.
With respect to the properties of the relative entropy\cite{Vedralrmp02}, it is quite easy to check that this measure satisfies the conditions of coherence measures.
In particular, there is a closed form solution that make it
easy to evaluate analytical expressions\cite{Baumgratz13}. For Hilbert space $\mathcal{H}$ with fixing the basis $\{|i\rangle\}_{i=1}^{d}$, we denote
\begin{equation}\label{eq:rho}
\rho=\sum_{i,i^\prime}\rho_{i,i^\prime}|i\rangle\langle i^\prime|
\end{equation}
and denote $\rho_{\mathrm{diag}}=\sum_i\rho_{ii}|i\rangle\langle i|$. By using the properties of relative entropy, it is easy to obtain
\begin{equation}\label{eq:rec1}
C_{\mathrm{RE}}(\rho)=S(\rho_{\mathrm{diag}})-S(\rho),
\end{equation}
where $S(\rho)=-\mathrm{Tr}\rho\log_2\rho$ is the von Neumann entropy\cite{Nielsen}.
We remark that the incoherent state $\rho_{\mathrm{diag}}$ is generated by removing all the off-diagonal elements and leaving the diagonal elements in density matrix or density operator $\rho$~(\ref{eq:rho}). This operation is called completely decohering, or completely dephasing channel\cite{Horodecki05}, we then denote
\begin{equation}\label{incoherent operation}
\rho_{\mathrm{diag}}=\Pi(\rho)=\sum_i^d\Pi_i\rho \Pi_i,
\end{equation}
where $\Pi_i=|i\rangle\langle i|$ are one-dimensional projectors, and $\sum_i\Pi_i=I_{\mathcal{H}}$, $I_{\mathcal{H}}$ is identity operator on Hilbert space $\mathcal{H}$. Thus, we claim that the coherence contained in quantum state is equal to the increased entropy caused by the completely decohering.
In addition, some basic properties have given\cite{Baumgratz13}. For example, we can obtain
\begin{equation}\label{rec_bound_1}
C_{\mathrm{RE}}(\rho)\leq S(\rho_{\mathrm{diag}})\leq\log d.
\end{equation}
Note that $C_{\mathrm{RE}}(\rho)=S(\rho_{\mathrm{diag}})$ if and only if the quantum state $\rho$ is a pure state.
In particular, if there exists pure states such that $C_{\mathrm{RE}}(\rho)=\log d$, these pure states are called maximally coherent states. Baumgratz \emph{et al.} have defined a maximally coherent state \cite{Baumgratz13}, which takes the form
\begin{equation}
|\psi\rangle=\frac{1}{\sqrt{d}}\sum_{i=1}^d|i\rangle.
\end{equation}

\noindent
{\bf Uncertainty-like relation between the coherence and entanglement.}
Interestingly, if one combines Eq.~(\ref{eq:rec1}) with Eq.~(\ref{rec_bound_1}), we can obtain a new tight bound of the relative entropy of coherence,
\begin{equation}
C_{\mathrm{RE}}(\rho)\leq I(\rho),
\end{equation}
where $I(\rho):=\log_2d-S(\rho)$ is the information function, which has an operational meanings: it is the number of pure qubits one can draw from many copies of the state $\rho$~\cite{Horodecki05}. By using a straightforward algebraic calculation, we can obtain an interesting ``uncertainty relation" between the coherence and the entropy of quantum system, namely,
\begin{equation}\label{uncertainty_relation}
C_{\mathrm{RE}}(\rho)+S(\rho)\leq \log_2d.
\end{equation}
This shows that the sum of the entropy of the quantum system and the amount of the coherence of quantum system is always smaller than a given fixed value: the larger $S(\rho)$, the smaller $C_{\mathrm{RE}}(\rho)$. In particular, when $\rho$ is the maximally mixed state,
then no coherence exists in the quantum system. But in another way the larger $C_{\mathrm{RE}}(\rho)$, the smaller $S(\rho)$.
Then, we can claim that if the quantum system is entangled with the outside world, then the coherence of the system may decay.

Next, we will discuss the relations between the coherence and entanglement in the bipartite quantum system.
Consider the bipartite quantum system in a composite Hilbert space $\mathcal{H}^{AB}=\mathcal{H}^A\otimes\mathcal{H}^B$, without loss of generality, we henceforth take $d=d_A=d_B$, where $d^A$ and $d^B$ are the dimensions of the quantum systems $A$ and $B$, which could be shared between two parties, Alice and Bob, respectively.
Let $\{|i\rangle^A\}_{i=1}^{d}$ and $\{|j\rangle^B\}_{j=1}^{d}$ be the orthogonal basis for the Hilbert space $\mathcal{H}^A$ and $\mathcal{H}^B$, respectively. Assume that a maximally coherent state of the bipartite quantum system is of the form
\begin{equation}\label{eq:mcs_bi}
|\psi\rangle^{AB}=\frac{1}{d}\sum_{i,j=1}^d|i\rangle^A|j\rangle^B.
\end{equation}
It is easy to verify that this state is a product state, i.e.,
\begin{equation}\label{eq:mcs_product}
|\psi\rangle^{AB}=\frac{1}{\sqrt{d}}\sum_{i=1}^d|i\rangle^A\otimes \frac{1}{\sqrt{d}}\sum_{i=1}^d|i\rangle^B.
\end{equation}
But that is not all the maximally coherent states can do, there is even a class of the maximally coherent states, they are also probably maximally entangled states. This shows that the maximally coherent state may be the maximally entangled state, or may be product state. This is because that the measure of the coherence depends on the choice of the basis, but the entangled property is not so. This also implies that though two states are both the maximally coherent states, their reduced states are entirely different. For the maximally entangled state with maximally coherent, its reduced states are completely mixed states, which does not exist the coherence. We give an example to illustrate the results as following.

\vspace{.5cm}
\noindent
{\bf Example 1} Consider two-qubit system with the basis $\{|00\rangle,|01\rangle,|10\rangle,|11\rangle\}$, and the relative entropy of coherence depends on this basis. Suppose that
\begin{equation}\label{eq:mes_max_1}
|\psi_{1}\rangle:=\frac{1}{2}(|00\rangle+|01\rangle-|10\rangle+|11\rangle).
\end{equation}
Obviously, we have that $C_{\mathrm{RE}}(|\psi_{1}\rangle)=2$. But at the same time, this state is also rewritten by
\begin{equation}\label{eq:mes_max_2}
|\psi_{1}^\prime\rangle=\frac{1}{\sqrt{2}}(|0\rangle|+\rangle+|1\rangle|-\rangle),
\end{equation}
where $|+\rangle=\frac{1}{\sqrt{2}}(|0\rangle+|1\rangle)$ and $|-\rangle=\frac{1}{\sqrt{2}}(|0\rangle-|1\rangle)$ are the maximally coherent states in one-qubit system.
Based on entanglement theory, we easily know that the state~(\ref{eq:mes_max_2}) is also a maximally entangled state. In addition,
it is generally known that Bell states are the maximally entangled states, one of them is
\begin{equation}\label{eq:mes_bi}
|\psi_{2}\rangle=\frac{1}{\sqrt{2}}\left(|00\rangle+|11\rangle\right).
\end{equation}
Obviously, it is not maximally coherent state. We easily give another maximally coherent state
\begin{equation}\label{eq:mcs_2qubit_1}
|\psi_{3}\rangle=\frac{1}{2}(|00\rangle+|01\rangle+|10\rangle+|11\rangle).
\end{equation}
This state is a product state, which is of the form
\begin{equation}\label{eq:mcs_2qubit_2}
|\psi_{3}^\prime\rangle=|+\rangle\otimes |+\rangle.
\end{equation}

Let $|\phi\rangle^{AB}=\sum_i\lambda_i|i\rangle^A|i\rangle^B$ be a bipartite entangled state (its Schmidt number is strictly greater than one) with respect to the basis in which the coherence is quantified. Then, the entanglement and the coherence are equal to the entropy of the subsystem, we have
\begin{equation}
E(|\phi\rangle^{AB})=C_{\mathrm{RE}}(|\phi\rangle^{AB})=S(\rho^A).
\end{equation}
Here, entanglement measure $E$ is any of distillation entanglement $E_D$\cite{Bennett96}, relative entropy of entanglement $E_{\mathrm{RE}}$\cite{Vedral97} and entanglement of formation $E_F$\cite{Wootters98}. They are upper bound on the entropy of subsystem and satisfy the inequality\cite{Horodecki00}
\begin{equation}
E_D(\rho^{AB})\leq E_{\mathrm{RE}}(\rho^{AB})\leq E_F(\rho^{AB})\leq \min\{S(\rho^A), S(\rho^B)\}.
\end{equation}
Then, we substitute this inequality into the uncertainty relation Eq.~(\ref{uncertainty_relation}) arriving at the following result.

\vspace{.5cm}
\noindent
{\bf Theorem 1} Given a quantum state $\rho^{AB}$ on the Hilbert space $\mathcal{H}^{AB}$, we have
\begin{equation}
E(\rho^{AB})+C_{\mathrm{RE}}(\rho^A)\leq \log_2 d^A.
\end{equation}
This inequality shows that the larger the coherence of subsystem, the less entanglement between two the subsystems. In other words, the system $A$ is already as entangled as it can possibly be with the other system $B$, then itself coherence would pay for their entangled behavior.
In analogy, if one builds quantum computer, to realize the purpose of computation, it is made clear that quantum computer has to be well isolated in order to retain its quantum coherence (or quantum properties). On the other hand, if one want to perform quantum information processing in term of the resource of entanglement, we expect to use maximally entangled state,
in this case, any information can not be obtained by local operation, for example in superdense coding and teleportation.

At the end of this section, we give another new property of the relative entropy of coherence.
Based on the additivity of the von Neumann entropy, we obtain that the relative entropy of coherence is additive,
\begin{equation}
C_{\mathrm{RE}}(\rho^{A}\otimes \rho^{B})=C_{\mathrm{RE}}(\rho^{A})+C_{\mathrm{RE}}(\rho^{B}).
\end{equation}
By using the properties of relative entropy, one
can show that the relative entropy of coherence satisfies the super-additivity.
Let $\Pi^A$ and $\Pi^B$ be two the completely dephasing operations on the subsystems $A$ and $B$, respectively. We denote $\Pi^{AB}=\Pi^A\otimes\Pi^B$, applying it on quantum state $\rho^{AB}$, we obtain the classical state $\rho_{\mathrm{diag}}^{AB}=\Pi^{AB}(\rho^{AB})$.
Since quantum operations never increase relative entropy, we then have
\begin{equation}
S(\Pi^{AB}(\rho^{AB})||\Pi^{AB}(\rho^A\otimes \rho^B))\leq S(\rho^{AB}||\rho^A\otimes\rho^B).
\end{equation}
Thus, we obtain the super-additivity inequality of the relative entropy of coherence,
\begin{equation}
C_{\mathrm{RE}}(\rho^{AB})\geq C_{\mathrm{RE}}(\rho^{A})+C_{\mathrm{RE}}(\rho^{B}).
\end{equation}
Obviously, for the maximally coherent state~(\ref{eq:mcs_bi}), the equality holds. From this relation, we know that the coherence contained in the bipartite quantum system is greater than the sum of the coherence of the local subsystems.

\vspace{.5cm}
\noindent
{\bf Relations between quantum coherence and quantum correlations.} We know that there are two different measures of quantum correlations via the different physical background, i.e., quantum discord and quantum deficit.
To better understand our results, let us give the formal definitions of the quantum discord and
one-way quantum deficit. For a bipartite quantum state $\rho^{AB}$, quantum discord is originally defined by the difference of two inequivalent
expressions for the mutual information via local von Neumann measurements\cite{Ollivier02},
\begin{equation}
\delta^{\rightarrow}(\rho^{AB}):=\min_{\{\Pi_i^A\}}(\mathcal{I}(\rho^{AB})-\mathcal{I}(\sum_i\Pi_i^A\rho^{AB}\Pi_i^A)).
\end{equation}
where the minimum is taken over all local von Neumann measurements on the subsystem $A$. Here $\mathcal{I}(\rho^{XY})=S(\rho^X)+S(\rho^Y)-S(\rho^{XY})$ is the mutual information\cite{Nielsen}.
Quantum deficit is originally defined by the difference the amount of extractable work for the global system and the local subsystems\cite{Oppenheim02}. In this paper, we only allow one-way classical communication from $A$ to $B$ by performed von Neumann measurements on the local system $A$, then the one-way quantum deficit\cite{Horodecki05} is defined as
\begin{equation}
\Delta^{\rightarrow}(\rho^{AB}):=\min_{\{\Pi_i^A\}}S(\rho^{AB}||\sum_i\Pi_i^A\rho^{AB}\Pi_i^A),
\end{equation}
where the minimum is taken over all local von Neumann measurements on the subsystem $A$.
Quantum discord and the one-way quantum deficit are nonnegative, and are equal to zero on classical-quantum states only. Horodecki \emph{et al.} have obtained that the one-way quantum deficit is an upper bound of quantum discord\cite{Horodecki05}, namely,
\begin{equation}
\delta^{\rightarrow}(\rho^{AB})\leqslant\Delta^{\rightarrow}(\rho^{AB}).
\end{equation}

In the following we will present some differences between them.
In general, we can always write
$\rho^{AB}=\sum_{i,i^\prime,j,j^\prime}\rho_{i,i^\prime,j,j^\prime}|i\rangle^A\langle i^\prime|\otimes |j\rangle^B\langle j^\prime|$ with fixed basis $\{|i\rangle^A|j\rangle^B\}_{i,j=1}^{d}$ for the bipartite quantum system, and
$\rho^A=\sum_{i,i^\prime}\rho_{i,i^\prime}|i\rangle^A\langle i^\prime|$ and $\rho^B=\sum_{j,j^\prime}\rho_{j,j^\prime}|j\rangle^B\langle j^\prime|$ are the reduced density operators or the reduced states for each party.
To extract information contained in the state, Alice can perform the measurement $\Pi$~(\ref{incoherent operation}) on her party, then the quantum state $\rho^{AB}$ becomes
\begin{equation}
\tilde{\rho}^{AB}=\sum_{i,j,j^\prime}\rho_{i,i,j,j^\prime}|i\rangle^A\langle i|\otimes |j\rangle^B\langle j^\prime|.
\end{equation}
and the reduced state $\rho^{A}$ becomes $\tilde{\rho}^A=\sum_i\rho_{ii}|i\rangle\langle i|$, but the reduced state $\rho^B$ does not change.
This shows that local measurement removes the coherent elements in the reduced state,
but it also destroys the quantum correlations between the parties $A$ and $B$. The post-measurement state $\tilde{\rho}^{AB}$ can be also written as
\begin{equation}\label{diagonal blocks}
\tilde{\rho}^{AB}=\sum_{i}p_i|i\rangle^A\langle i|\otimes \rho^B_i,
\end{equation}
where $\rho^B_i=\sum_{j,j^\prime}\rho_{i,i,j,j^\prime}|j\rangle^B\langle j^\prime|/\mathrm{Tr}(\sum_{j,j^\prime}\rho_{i,i,j,j^\prime}|j\rangle^B\langle j^\prime|)$
is the remaining state of $B$ after obtaining the outcome $i$ on $A$ with the probability $p_i=\mathrm{Tr}(\sum_{j,j^\prime}\rho_{i,i,j,j^\prime}|j\rangle^A\langle j^\prime|)$. It is also easy to check that $p_i=\sum_{j}\rho_{i,i,j,j}=\rho_{ii}$.
By the local measurement $\Pi$, Alice can extract information which can be given by the mutual information about the classical-quantum state $\tilde{\rho}^{AB}$,
\begin{equation}
\mathcal{J}^{\rightarrow}(\rho^{AB}|\Pi):=\mathcal{I}(\tilde{\rho}^{AB}).
\end{equation}
The quantity $\mathcal{J}^{\rightarrow}(\rho^{AB}|\Pi)$ represents the amount of information gained about the subsystem $B$ by measuring the subsystem $A$.
We use the difference of mutual information before and after the measurement $\Pi$ to characterize the amount of quantum correlation in quantum state $\rho^{AB}$,
\begin{equation}
\delta^{\rightarrow}(\rho^{AB}|\Pi)=\mathcal{I}(\rho^{AB})-\mathcal{I}(\tilde{\rho}^{AB}).
\end{equation}
The quantification $\delta^{\rightarrow}(\rho^{AB}|\Pi)$ is the discord-like quantity. Then, we can define the deficit-like quantity (full name, one-way quantum deficit-like) $\Delta^{\rightarrow}(\rho^{AB}|\Pi)$ with respect to the local measurement $\Pi$,
\begin{equation}
\Delta^{\rightarrow}(\rho^{AB}|\Pi)=S(\rho^{AB}||\tilde{\rho}^{AB}).
\end{equation}
More explicitly we have $\Delta^{\rightarrow}(\rho^{AB}|\Pi)=\Delta S_{AB}$, where $\Delta S_{AB}=S(\tilde{\rho}^{AB})-S(\rho^{AB})$
is the increased entropy produced by the local measurement on $A$.
After some algebraic manipulation,
we give firstly trade-off as follows
\begin{align}\label{eq:trade-off_1}
\delta^{\rightarrow}(\rho^{AB}|\Pi)+C_{\mathrm{RE}}(\rho^A)=\Delta^{\rightarrow}(\rho^{AB}|\Pi).
\end{align}
This shows that the increased entropy produced by the local measurement is equal to the sum between the quantum correlation destroyed by the local measurement and the relative entropy of coherence of the measured system. Note that the trade-off only holds with respect to the local measurement $\Pi$. But, we know that the discord and the deficit do not depend on the local measurement.
In the following, we will discuss the general case. If one optimizes the discord-like quantity and deficit-like quantity
over all the rank-1 projective measurements, then we can obtain the second trade-off relation between them as follows.

\vspace{.5cm}
\noindent
{\bf Theorem 2}
Given a quantum state $\rho^{AB}$ on the Hilbert space $\mathcal{H}^{AB}$, if $\delta^{\rightarrow}(\rho^{AB})>0$, then we have
\begin{equation}\label{eq:trade-off_2}
\delta^{\rightarrow}(\rho^{AB})+C_{\mathrm{RE}}(\rho^A)=\Delta^{\rightarrow}(\rho^{AB}).
\end{equation}
The proof is left to the Method.
This shows that the measures of quantum correlations are distinct from each other with respect to the different background, but this difference does not affect the inherent quantum correlation between the subsystems, this difference can be described exactly by the coherence of the measured system.
Note that the condition $\delta^{\rightarrow}(\rho^{AB})>0$ is necessary. If not, let us consider the state $|\psi_3\rangle$
 in the Example 1, we have that $\delta^{\rightarrow}(\rho^{AB})=\Delta^{\rightarrow}(\rho^{AB})=0$, but $C_{\mathrm{RE}}(\rho^A)=1$.

After the local measurements, we only obtain the diagonal blocks matrix~(\ref{diagonal blocks}).
That is to say, to obtain completely diagonal matrix with respect to the basis $\{|i\rangle^A|j\rangle^B\}$, we
must remove the all off-diagonal elements, and remain the diagonal elements in every the diagonal block matrix. For every block matrix $\rho^B_i$,
we can perform the similar operation~(\ref{incoherent operation}) in the previous section. After performing these operations, it follows that
\begin{equation}\label{eq:cc state}
\rho_{\mathrm{diag}}^{AB}=\sum_{i,j}\rho_{i,i,j,j}|i\rangle^A\langle i|\otimes |j\rangle^B\langle j|.
\end{equation}
Obviously, the state $\rho^{AB}$ has the same incoherent state as the classical-quantum state $\tilde{\rho}^{AB}$. Based on this fact, by using the approach of the proof of the Theorem 2, we then obtain the third trade-off relation,
\begin{equation}\label{eq:trade-off_3}
C_{\mathrm{RE}}(\rho^{AB})-C_{\mathrm{RE}}(\tilde{\rho}^{AB})=\Delta^{\rightarrow}(\rho^{AB}).
\end{equation}
Intuitively, the local measurement can lead to the decrease of the coherence in bipartite quantum system.
That is to say, quantum correlations in the bipartite quantum system is equal to the amount of the coherence lost
by the measurement on one of the subsystems.

\vspace{.5cm}
\noindent
{\large\bf Discussion}

\vspace{.1cm}
\noindent
We have obtained two new properties of the relative entropy of coherence, the one is that the relative entropy of coherence does not exceed the information function for a given quantum state, the other is the super-additivity. Based on the former, we have obtained an uncertainty-like relation between the coherence and the entropy of quantum system, i.e., the more the coherence, the less the entropy. We have obtained another uncertainty-like relation between the entanglement and the coherence of subsystem, i.e., the system is already as entangled as it can possibly be with the outside world, then the coherence itself would pay for their entangled behavior.
For any bipartite quantum system, by performing completely dephasing operation on the subsystem, we have obtained three trade-offs among the relative entropy of coherence, quantum discord-like
and one-way quantum deficit-like quantum correlations. Our results gave a clear quantitative analysis and operational connections between quantum coherence and quantum correlations in the bipartite quantum system.
We may focus further on the fascinating question whether one can find the relation between two-way quantum deficit and the relative entropy of coherence. It is also possible that all four concepts, thermodynamics, entanglement, quantum correlations and coherence,
can be understood in a unified framework. Those progresses may develop further the quantum information science.

\vspace{.5cm}
\noindent
{\large\bf Methods}

\vspace{.1cm}
\noindent
{\bf Proof of the Theorem 2 in the Main Text.}
Before proceeding with the proof of Theorem 2, a fact is that the relative entropy of coherence is
unitary invariant by using the different bases. For $d$-dimensional Hilbert space $\mathcal{H}$, we can take the basis $\{|i\rangle\}_{i=1}^{d}$, then the density operator upon it can be given by Eq.~(\ref{eq:rho}).
Under the unitary operator $U$, the density operator~(\ref{eq:rho}) become
\begin{equation}\label{eq:rho_A2}
\rho_U:=U\rho U^\dagger=\sum_{i,i^\prime}\rho_{i,i^\prime}U|i\rangle\langle i^\prime|U^\dagger=\sum_{i,i^\prime}\rho_{i,i^\prime}|\varphi_i\rangle\langle \varphi_{i^\prime}|,
\end{equation}
where $|\varphi_i\rangle=U|i\rangle$ for each $i$. Obviously, the density operators $\rho$ and $\rho_U$ have same the matrix elements under the bases $\{|i\rangle\}_{i=1}^{d}$ and $\{|\varphi_i\rangle\}_{i=1}^{d}$, respectively.
Then, we denote $C_{\mathrm{RE}}(\rho)$ as the measure of coherence under the basis $\{|i\rangle\}_{i=1}^{d}$, and denote $C_{\mathrm{RE}}(\rho_U)$ as the measure of coherence under the basis $\{|\varphi_i\rangle\}_{i=1}^{d}$, we obtain
\begin{equation}
C_{\mathrm{RE}}(\rho)=C_{\mathrm{RE}}(\rho_U).
\end{equation}
Then, we begin the proof of Theorem 2.
Let $\{|i\rangle^A|j\rangle^B\}$ be the orthogonal basis for the Hilbert space $\mathcal{H}^{AB}$, and the bipartite quantum state can be given by
\begin{equation}
\rho^{AB}=\sum_{i,i^\prime,j,j^\prime}\rho_{i,i^\prime,j,j^\prime}|i\rangle^A\langle i^\prime|\otimes |j\rangle^B\langle j^\prime|.
\end{equation}
Let $\{\Pi^{\delta}_i\}$ be an optimal projective measurement for quantum discord $\delta^{\rightarrow}(\rho^{AB})$. By using this measurement, we can define a new basis on the Hilbert space $\mathcal{H}^A$, denote $\Pi^{\delta}_i=|i\rangle_{\delta}\langle i|$. Without loss of generality, let $\{|i\rangle^{\delta}|j\rangle\}$ be the basis on the Hilbert space $\mathcal{H}^{AB}$, then there exists an unitary operator $U^A$ on $A$ such that
\begin{align}
\rho^{AB}_\delta=&(U^A_\delta\otimes I^B)\rho^{AB}(U^A_\delta\otimes I^B)^\dagger\nonumber\\
=&\sum_{i,i^\prime,j,j^\prime}\rho_{i,i^\prime,j,j^\prime}|i\rangle^A_\delta \langle i^\prime|\otimes |j\rangle^B\langle j^\prime|.
\end{align}
By using the properties of the discord and the deficit\cite{Modi12}, we know
\begin{align}\label{app_eq:discord1}
\delta^{\rightarrow}(\rho^{AB})=&\delta^{\rightarrow}(\rho^{AB}_\delta)=\delta^{\rightarrow}(\rho^{AB}_\delta|\Pi^\delta),\nonumber\\
\Delta^{\rightarrow}(\rho^{AB})=&\Delta^{\rightarrow}(\rho^{AB}_\delta)\leq \Delta^{\rightarrow}(\rho^{AB}_\delta|\Pi^\delta).
\end{align}
Using Eq.~(\ref{eq:trade-off_1}), under the basis $\{|i\rangle_{\delta}\}_{i=1}^{d}$, we have
 \begin{align}
\delta^{\rightarrow}(\rho^{AB}_\delta|\Pi^\delta)+C_{\mathrm{RE}}(\rho^A_\delta)=\Delta^{\rightarrow}(\rho^{AB}_\delta|\Pi^\delta).
\end{align}
Substituting Eqs.~(\ref{app_eq:discord1}) into this relation, we obtain
 \begin{equation}\label{app_eq:upper_deficit}
\delta^{\rightarrow}(\rho^{AB})+C_{\mathrm{RE}}(\rho^A_\delta)\geq\Delta^{\rightarrow}(\rho^{AB}).
\end{equation}
Similarly, let $\{\Pi^{\Delta}_i\}$ be an optimal projective measurement for the one-way quantum deficit $\Delta^{\rightarrow}(\rho^{AB})$. We can also define another basis on the Hilbert space $\mathcal{H}^A$, denote $\Pi^{\Delta}_i=|i\rangle_{\Delta}\langle i|$. Let $\{|i\rangle^{\Delta}|j\rangle\}$ be the basis on the Hilbert space $\mathcal{H}^{AB}$, then there exists an unitary operator $U^A_\Delta$ on $A$ such that
\begin{align}
\rho^{AB}_\Delta=&(U^A_\Delta\otimes I^B)\rho^{AB}(U^A_\Delta\otimes I^B)^\dagger\nonumber\\
=&\sum_{i,i^\prime,j,j^\prime}\rho_{i,i^\prime,j,j^\prime}|i\rangle^A_\Delta \langle i^\prime|\otimes |j\rangle^B\langle j^\prime|.
\end{align}
Naturally, we have the following relations
\begin{align}\label{app_eq:discord2}
\delta^{\rightarrow}(\rho^{AB})=&\delta^{\rightarrow}(\rho^{AB}_\Delta)\leq\delta^{\rightarrow}(\rho^{AB}_\Delta|\Pi^\Delta),\nonumber\\
\Delta^{\rightarrow}(\rho^{AB})=&\Delta^{\rightarrow}(\rho^{AB}_\Delta)= \Delta^{\rightarrow}(\rho^{AB}_\Delta|\Pi^\Delta).
\end{align}
Then, depending on the bases $\{|i\rangle_{\Delta}\}_{i=1}^{d}$, by using Eqs.(\ref{app_eq:discord2}), we have
 \begin{align}\label{app_eq:upper_deficit2}
\Delta^{\rightarrow}(\rho^{AB})
=&\delta^{\rightarrow}(\rho^{AB}_\Delta|\Pi^\Delta)+C_{\mathrm{RE}}(\rho^A_\Delta)\nonumber\\
\geq& \delta^{\rightarrow}(\rho^{AB}_\Delta)+C_{\mathrm{RE}}(\rho^A_\Delta)\nonumber\\
=&\delta^{\rightarrow}(\rho^{AB})+C_{\mathrm{RE}}(\rho^A_\Delta).
\end{align}
Combining Eq.~(\ref{app_eq:upper_deficit}) with Eq.~(\ref{app_eq:upper_deficit2}), we obtain the following relation
\begin{equation}\label{app_eq:discord_deficit_coherence}
\delta^{\rightarrow}(\rho^{AB})+C_{\mathrm{RE}}(\rho^A_{\delta})\geq \Delta^{\rightarrow}(\rho^{AB}) \geq\delta^{\rightarrow}(\rho^{AB})+C_{\mathrm{RE}}(\rho^A_{\Delta}).
\end{equation}
By using the fact in the previous, we have
\begin{equation}
C_{\mathrm{RE}}(\rho^A_\Delta)=C_{\mathrm{RE}}(\rho^A_\delta)=C_{\mathrm{RE}}(\rho^A).
 \end{equation}
 Substituting this relation into Eq.~(\ref{app_eq:discord_deficit_coherence}), we obtain
\begin{align}
\delta^{\rightarrow}(\rho^{AB})+C_{\mathrm{RE}}(\rho^A)=\Delta^{\rightarrow}(\rho^{AB}).
\end{align}
Thus, we get the desired result.

\vspace{.5cm}
\noindent
{\large \bf Acknowledgments}

\vspace{.01cm}
\noindent
The authors thank Prof. Y. Feng for helpful discussion. Z.J. Xi is supported by NSFC (Grant No. 61303009),  and Specialized Research Fund for the Doctoral Program of Higher Eduction (20130202120002), and Fundamental Research Funds for the Central Universities (GK201502004). Y.M. Li is supported by NSFC (Grant No. 11271237). H. Fan is supported by NSFC (Grant No.11175248).

\vspace{.5cm}
\noindent
{\large \bf Author contributions}

\vspace{.01cm}
\noindent
Z.X. contributed the idea. Z.X. and H.F. performed the calculations.
Y.L. checked the calculations. Z.X. wrote the main manuscript,  Y.L. and H.F. made an
improvement. All authors contributed to discussion and reviewed the manuscript.

\vspace{.5cm}
\noindent
{\large \bf Additional Information}

\vspace{.01cm}
\noindent
\textbf{Competing financial interests:} The authors declare no competing financial interests.
\end{document}